# Sustainable Aviation Fuels: Opportunities, Alternatives and Challenges for Decarbonizing the Aviation Industry and Foster the Renewable Chemicals

**Wesley Bonicontro Ambrosio[a,b], Bruna Araújo de Sousa[a,c], João Marcos Kanieski [c,d], Priscila Marchiori[e], Gustavo Mockaitis[a*]**

[a]Interdisciplinary Research Group on Biotechnology Applied to the Agriculture and the Environment, School of Agricultural Engineering, University of Campinas (GBMA/FEAGRI/UNICAMP), Campinas, 13083-875, Brazil

[b]Braskem S/A- Avenida Wagner Samara, N°1280, Bairro Cascata, 13140-031, Paulínia, SP, Brazil

[c]Interinstitutional Graduate Program in Bioenergy (USP/UNICAMP/UNESP) – 330 Cora Coralina Street, Cidade Universitária, SP, Campinas, CEP 13.083-896, Brazil

[d]Faculdade de Engenharia Elétrica e de Computação (FEEC), Universidade Estadual de Campinas (UNICAMP), Av. Albert Einstein, Nº 400, 13083-852 Campinas, SP, Brasil

[e]Faculdade de Engenharia de Alimentos (FEA), Universidade Estadual de Campinas (UNICAMP), Rua Monteiro Lobato, n.80, 13083-862 Campinas, SP, Brasil

* Corresponding author: Bruna Araújo de Sousa - E-mail address: b247331@dac.unicamp.br

---

## Abstract

Sustainable Aviation Fuels (SAF) are pivotal in the global effort to decarbonize the aviation sector and meet greenhouse gas (GHG) reduction targets established by international frameworks such as CORSIA and Brazil's ProBioQAV. This study evaluates SAF's potential to reduce lifecycle carbon emissions by up to 80% while being compatible with existing aviation infrastructure. Through bibliometric analysis, scenario evaluation, legal and regulatory framework analysis and economic modeling, the research examines two key SAF production technologies: Hydroprocessed Esters and Fatty Acids Synthetic Paraffinic Kerosene (HEFA-SPK) and Alcohol-to-Jet (ATJ) pathways in the Brazilian context.

The findings reveal significant economic challenges, particularly high feedstock and production costs, which hinder SAF's competitiveness with fossil fuels at recent and current market prices in Brazil, leading to the analysis of potential incentives and commercial conditions aiming to increase economic attractiveness of SAF production. Based on interviews with relevant stakeholders and decision makers in the industry, scenarios incorporating tax incentives, carbon credits, capital grants, and premium pricing for SAF and its biogenic by-products demonstrate that combined policy interventions and commercial arrangements, along with a regulated Carbon Market are essential for SAF economic viability. The study highlights the critical role of supportive public policies, infrastructure development, and collaboration among stakeholders in scaling SAF production and achieving mandated decarbonization goals.

Future research is suggested to look at regional assessments of feedstock availability, supply chain logistics, and global market eligibility. Additionally, the broader economic impacts of SAF and its biogenic by-products on value chains such as agriculture and renewable chemical and petrochemical industries, as well as societal benefits including job creation and GDP growth, indicate potential further investigation. Exploration of innovative policy frameworks, such as regulatory sandboxes and funding schemes are recommended to foster SAF industry growth and mitigate investment risks. This research provides insights for guiding public policy and private investment to support the transition to sustainable aviation in Brazil and beyond.

## 1. Introduction

The consumption of aviation fuels derived from fossil sources has been growing annually among commercial airlines. Aviation accounts for about 2.8% of global greenhouse gas (GHG) emissions. With the increase in mandates and policies that promote the use of renewable fuels, the production of sustainable aviation fuels (SAF) – particularly those from non-petroleum-based sources – will be essential to decarbonize the sector and reduce net GHG emissions, aiming to achieve net-zero carbon emissions by 2050 [1], [2], [3].

Research indicates that SAF can reduce carbon emissions over its life cycle by up to 80%, as well as decrease emissions of other pollutants, such as soot, sulfur oxides, and ultrafine particulate matter [3]. The use of SAF also helps to mitigate the climate impact of condensation trails left by aircraft. However, the allowable blending rates for SAF, depending on the certified processes, are limited, ranging from 5% to 50%, depending on the type of conversion process used [3], [4].

SAF is a viable substitute for fossil aviation fuels, designed to closely resemble the physical and chemical properties of conventional fossil fuels. Chemically, SAF is primarily a blend of aliphatic hydrocarbons, similar to conventional aviation kerosene, including paraffins, iso-paraffins, and some olefins. The term "drop-in" implies that SAF can be used without requiring modifications to engines or their components. While drop-in fuels can have a similar composition to conventional fossil fuels, they must be blended with conventional fuels to achieve the desired performance [13]. The aromatic content in fuel can serve as an indicator of particulate matter (PM) emissions when burned, thus, lower aromatic levels result in reduced PM emissions. The type of feedstock used in production will determine the composition, intermediates, strengths, and challenges of each technology and feedstock, which are evaluated under ASTM D4054 [40] to verify their equivalence to conventional aviation fuel.

SAF (Sustainable Aviation Fuel) can be described as a "drop-in" replacement for traditional aviation fuel, being composed of synthetic fuels or biofuels that meet sustainability criteria. For the fuel to be considered SAF, it must meet some criteria: (1) contribute to the reduction of net greenhouse gas (GHG) emissions and carbon footprint throughout its life cycle, (2) limit the use of fresh water during the production process, and (3) must not require deforestation or compete with areas intended for food and crop production [2], [5].

Given the growing focus on sustainable alternatives for the aviation sector, this article proposes a literature review on sustainable aviation fuels (SAF) as a way to guide future investments in scientific, technological and business research and development. The relevance of SAF lies in its ability to significantly reduce greenhouse gas (GHG) emissions and contribute to the sector's decarbonization goals by 2050. However, the development and implementation of these fuels still face technical, economic, and regulatory barriers, as well as the need for adequate infrastructure for large-scale support. This article is also justified by the need to identify, recommend and review public policies that promote the growth of the SAF industry, while evaluating the infrastructure necessary for its production and distribution. Additionally, it aims to point out gaps in the existing literature and outline key questions that need to be answered, providing a solid foundation for directing future research and investment efforts.

## 2. Methodology

### 2.1 Bibliometric Research

The research was designed on two fronts: the first was SAF + politics + economy and the second was SAF + technologies + production. The following were used for data collection: Scopus, Web of Science and Science Direct. To refine the search, the following descriptors were used for the first search: ("bio kerosene" OR "biojet" OR "SAF") AND (“economic" )

NOT ( "catalytic" OR "oxidation" OR "production" OR "neural" OR "cell" OR "Systems Approach Framework"), and for the second search: ("bio kerosene" OR "biojet" OR "SAF") AND "production" AND ("HEFA" OR "Alcohol-to-Jet" OR "Fischer-Tropsch"). For the inclusion criteria, only "research articles" were selected for refinement and journals that did not include the "energy and sustainability" criteria were removed, with the first research being more focused on studies of political/economic/regulatory interest and, for the second research, the focus was related to SAF production technologies.

The data obtained from the analysis in question, together with the inclusion and exclusion criteria, were obtained for subsequent bibliometric analysis. The data were collected in *bib.txt* format and analyzed using two software programs, VOSviewer and the bibliometrix package in R language, with the removal of duplicate works. The analyses included authors, co-authors, countries, institutions, journal sources and keywords, obtaining at the end network maps and tables with the correlations between the articles and the metrics for the researched topics.

## 2.2 Sensitiveness Analysis

This study is based on the scenario analysis proposed by EPE (Empresa de Pesquisa Energética - Energetic Research Enterprise), a Brazilian governmental institution, forced to provide data to support policy-making decisions in the energy sector in Brazil. Because each route to SAF has a different carbon footprint, the total volume of SAF required to meet legal targets varies according to different route adoption scenarios. Announced projects consider opportunities using traditional and alternative raw materials such as ethanol, fresh and reused vegetable oils, tallow and others. Future scenarios also consider the use of agriculture biomass and residues, among others.

Scenario I: This scenario considers only the projects announced: BBF | HEFA production in 2026: 250 thousand m³/year raw material: palm, soybeans and corn. Acelen | HEFA

production in 2027: 500 thousand m³/year raw material: soybeans, corn and macaúba palm oil (in the future). Petrobras | HEFA production in 2029: 350 thousand m³/year raw material: soybean tallow and beef.

From 2027 to 2037, projects achieve an average of 38% of the emission reduction targets set by CORSIA (Carbon Offsetting and Reduction Scheme for International Aviation) and National Sustainable Aviation Fuel Program - ProBioQAV. Considering ProBioQAV alone, the announced projects are sufficient to meet the established targets by 2037 if all projects meet the announced targets.

Scenario II: Considers the use of traditional feedstocks, such as soybean oil, first-generation sugarcane, and corn ethanol, and also the use of alternative feedstocks, such as cellulosic ethanol and agave, and macauba oil. The combination of conversion processes and raw material will depend on the evaluation of various factors, for example, raw material availability, logistics, costs, environmental aspects among others. By 2037, SAF production is expected to range from 3.7 to 8 million m³/year, depending on the conversion processes chosen, this production range includes the announced projects. SAF can represent between 36% and 78% of the volumetric demand for aviation fuel.

Scenario III: This scenario considers the possibility of using organic residues from large-scale agro-industrial activities: Bovine tallow, Sugarcane residues and forest residues (*Eucalyptus spp.*). The use of organic waste is attractive due to the low cost of raw material acquisition and low carbon intensity. This trajectory considers the level of use of available waste. If the full utilization potential were realized, it would be possible to meet 82% of the 2037 emissions reduction targets using only residual biomass.

Considering the assessment made by EPE and looking at scenario II proposed in the study the authors developed a simplified economic analysis to compare HEFA-SPK and ETJ routes, aiming to assess relative competitiveness and to identify potential levers that could incentivize the investment required in the Brazilian industry to meet the GHG reduction goals. The method is presented in item **7 - Economic Evaluation of HEFA-SPK and ETJ routes**.

## 3. Results and Discussion

### 3.1 Bibliometric Research

According to the images of graphs on SAF networks (Figures 1 and 2), we can highlight some keywords that fill the main interconnections, in which they delimit the paths of studies that are in focus. The first of them linked to "emission" and "environmental impact", demonstrating the possibilities that the use of alternative aviation fuel SAF can help reduce emissions and environmental impact with its use [6].

Other terms that are connected are "conversion" and "technological/economic analysis" are in attention due to the need to analyze the routes for the production of SAF and its technical economic feasibility (Figures 1, 2, 3). The specific technologies in evidence are the Fischer-Tropsch synthesis and oligomerization, the first method of SAF production in accordance with ASTM specifications, consists of a process of production of syngas, following the generation of synthetic crude oil and refining of this material, which may have organic waste as raw materials, forestry and urban [5], [7]. The second technology/process, oligomerization is a step of the Alcohol-to-jet process that uses starches, sugars and cellulosic biomass as raw materials, the process for obtaining SAF is through dehydration, oligomerization and hydrogenation of substances containing alcohols. These two technologies are essential for the transformation of syngas and alcohols among other precursors of aviation fuels, with the conversion of biomass materials or waste into liquid fuels with low potential for environmental impact [8], [9].

In the three graphs, it is observed the need for an economic analysis that goes beyond the definition of routes, also covering the choice of raw material to ensure efficiency and lower cost in the long term. The terms "vegetable oil", "microalgae" and "lignin" stand out for their high volume of studies and are related to emerging technologies, such as Fischer-Tropsch and HEFA. In particular, the use of municipal and forest solid waste has been the focus of attention and has already been applied on an industrial scale since 2009, by the Shell company. Another promising raw material involves oils and fats, which can come from vegetable oils and/or microalgae. These materials are processed via HEFA for the production of SAF, a process consolidated at an industrial level since 2011, implemented by companies such as Honeywell and Neste Oil. In addition to reducing the carbon footprint with the use of SAF, the use of industrial and urban waste contributes even more to this reduction. In summary, there is a diversity of raw materials and technologies with high potential for sustainable and efficient use [7], [10].

Visualizing the geopolitical perspectives of SAF production, Brazil shows relevant influence in South America and Germany, in Europe. Both regions present policies and incentives in the implementation of SAF production technology. Depending on each country, the economic bloc may favor certain types of technological routes, which are interconnected with its energy and sustainability policies. National decarbonization policies and economic challenges are terms that appear in the word cloud, suggesting that SAF routes are not only dependent on technological feasibility, but also on the regulatory and incentive layer to make them competitive with fossil fuels [5], [9]. It is also possible to identify the term "blends" that demonstrate that the transition to SAF will be gradual, to seek the adaptation of the aviation sector [11], [12].

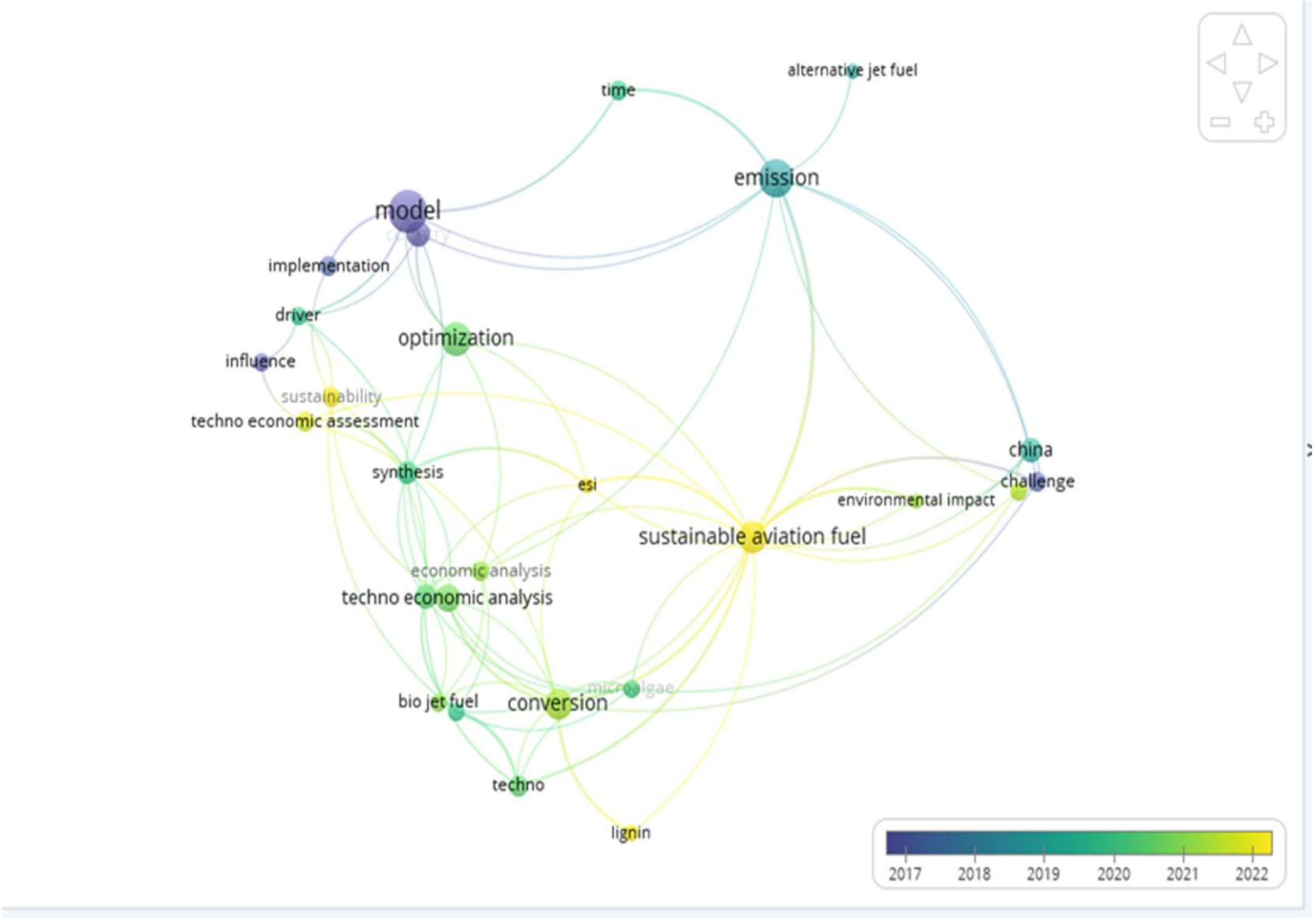

**Figure 1** - Combining Research 1 and 2: Improved Image Clouds with 10 Repetitions, Using Only Article Titles, Removing Repeated Words and Those Not Related to the Energy Topic**.**

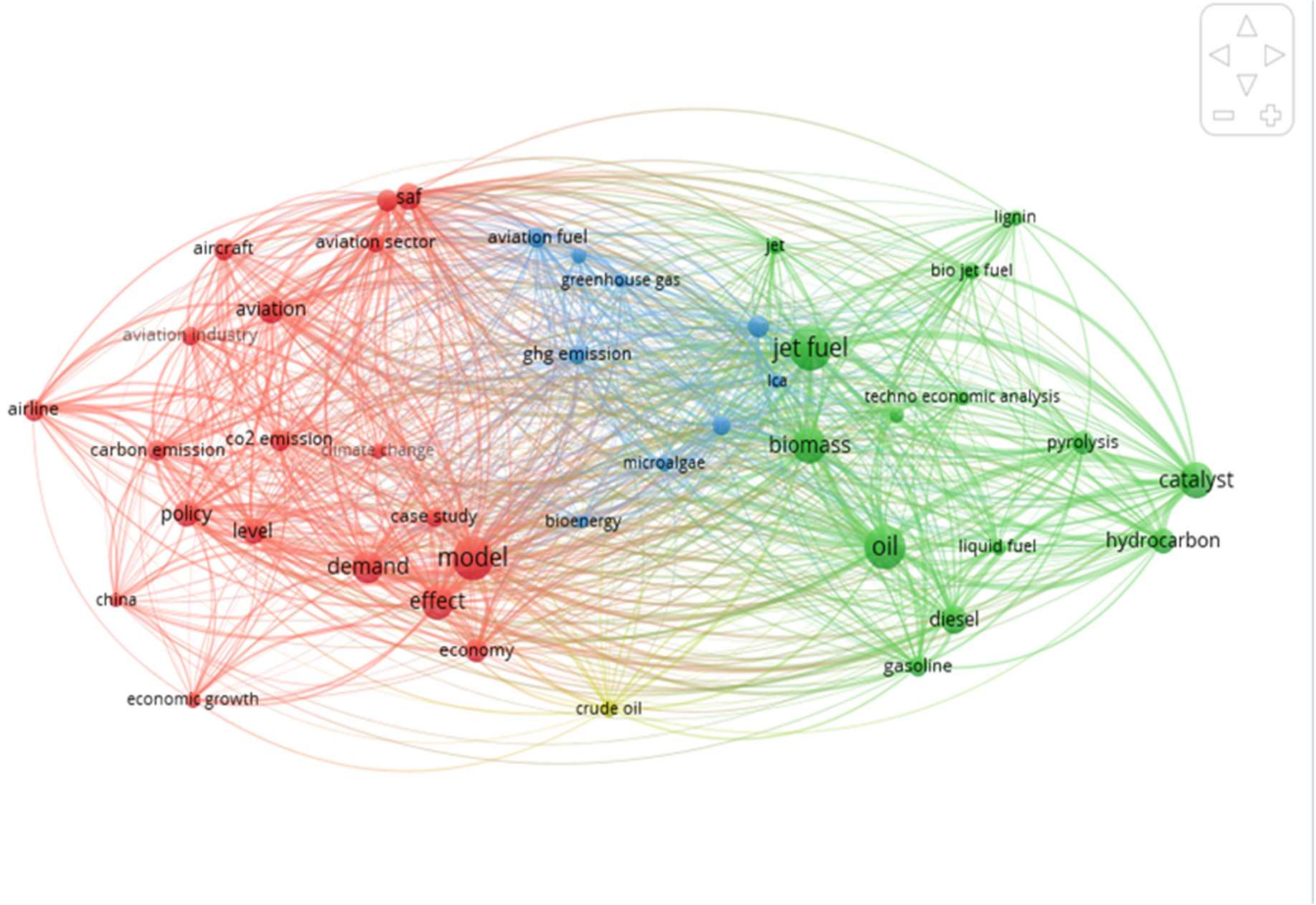

**Figure 2. Word cloud with Search1: ("jet fuel" OR "biojet") AND ("economic")**

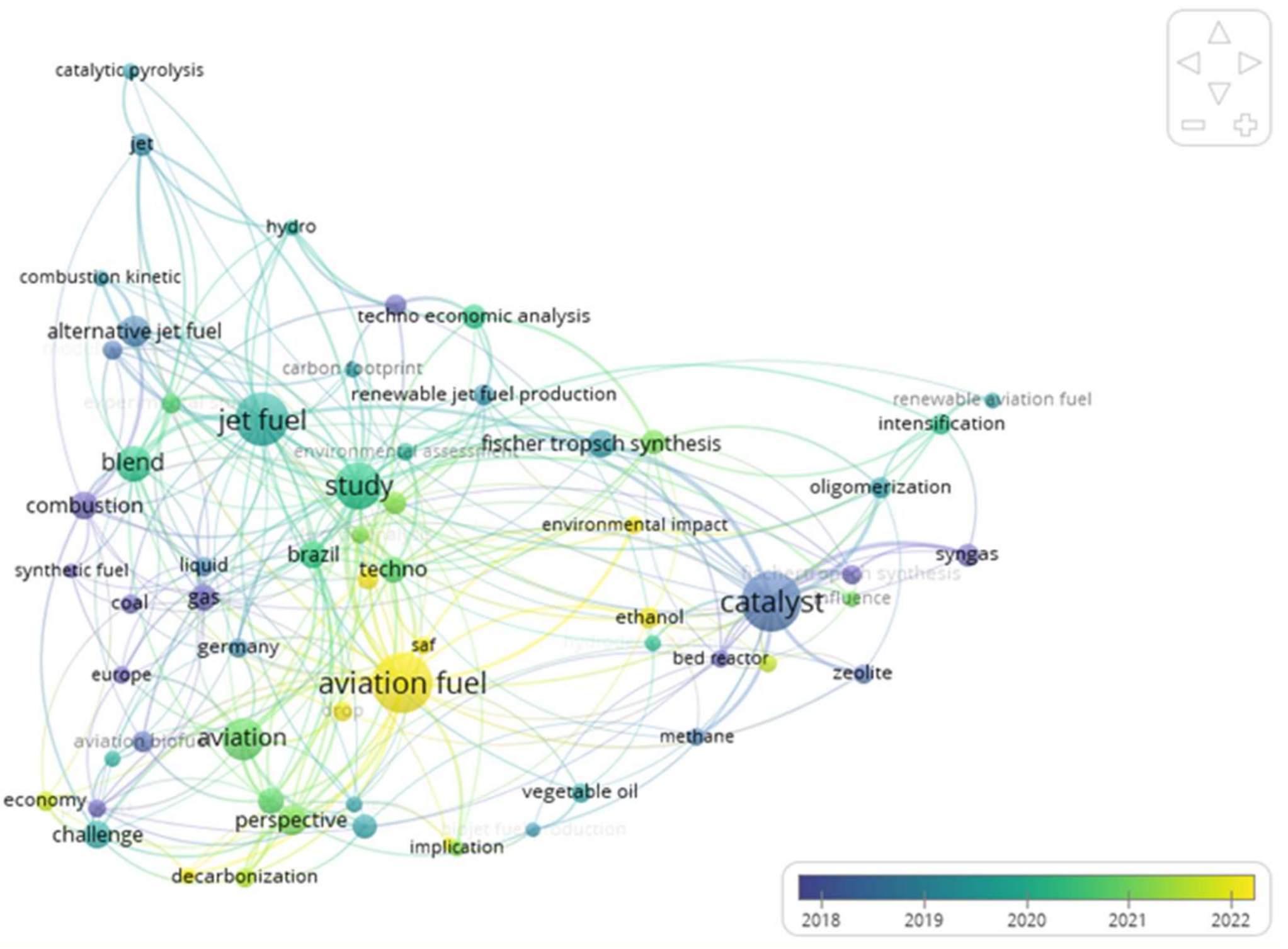

**Figure 3 - Word cloud with the combination of technology searches (HEFA, Fischer, Alcool-to-jet)**

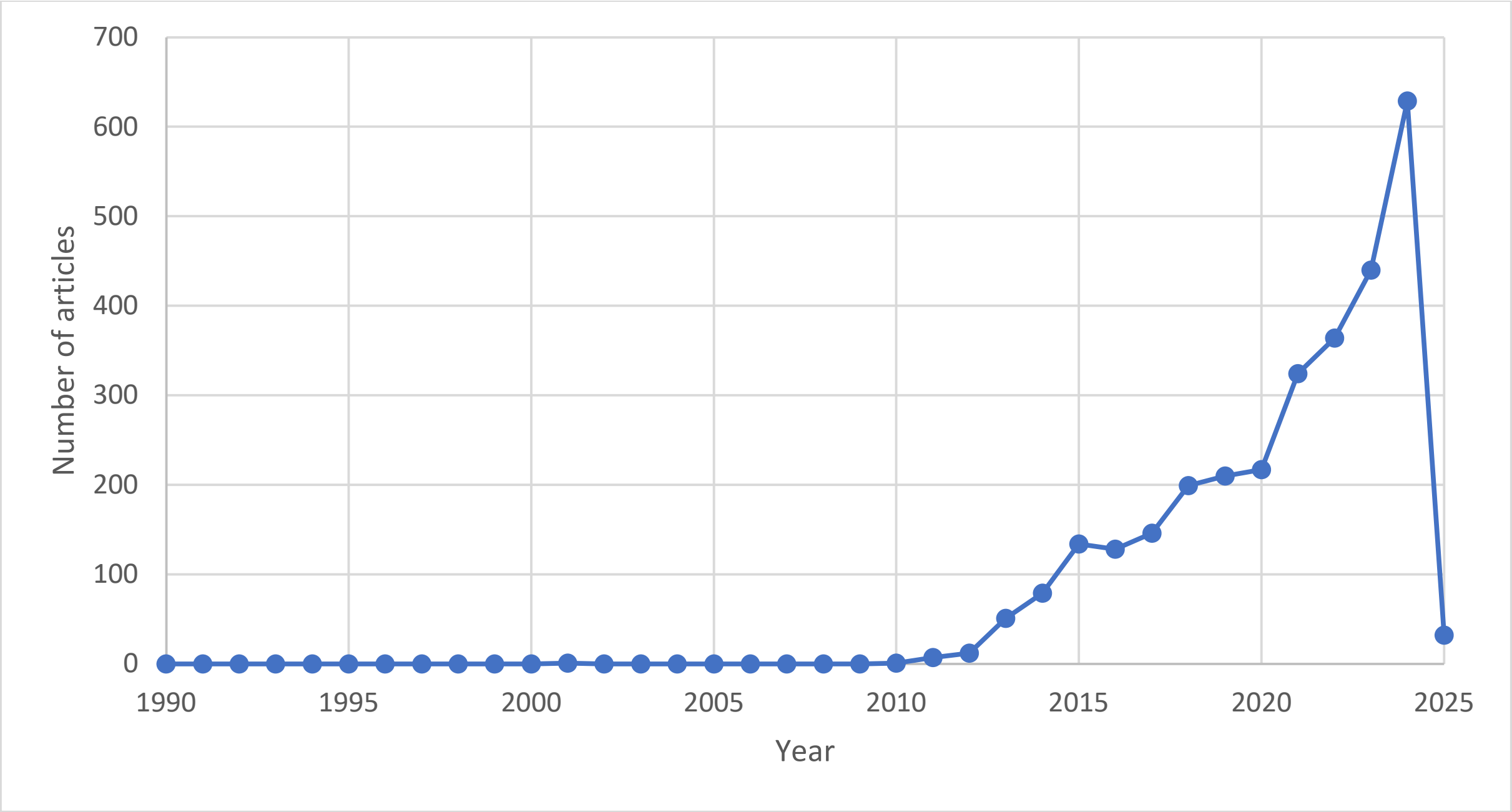

**Figure 4 - Annual scientific production related to "SAF" and "bio kerosene" and production. The arrows represent important articles, Arrow A: 1 article; Arrow B: 8 articles; Arrow C: Beginning of the accelerated increase in production.**

The graph shows an exponential growth curve related to the number of scientific publications (research articles) related to aviation fuels over the last few years. Initially in 1986 (data not shown in the graph), Lander *et al.*[35] demonstrated one of the first alternatives to the use of kerosene refined from petroleum, the use of kerosene produced from oil shale, a sedimentary rock rich in kerogen [35]. The compound, when subjected to pyrolysis processes and subsequent refining, enables the generation of aviation kerosene. However, despite representing a technological innovation for the time, this method still depends on a fossil source, which limits its positive sustainable impact. In subsequent years, research efforts focused on improving the processes of refining oil and oil shale for the production of aviation fuel.

With the signing of the Kyoto Protocol in 1997 [36], during the 3rd Conference of the Parties to the United Nations Framework Convention on Climate Change (UNFCCC), a global target for reducing greenhouse gas (GHG) emissions was established. This political milestone encouraged increased research focused on developing fuels from renewable sources and limiting $CO_2$ emissions, with the race for alternative fuels beginning in 2000 (point A - Figure 4). From that moment on, studies on technologies such as pyrolysis, biogas, ethanol production and biodiesel have intensified, aiming to find sustainable alternatives for the most diverse human needs for fossil fuels.

In 2001, R. O. Dunn published a paper entitled "Alternative Jet Fuels from Vegetable Oils" [37], in which he studied the use of biodiesel derived from vegetable oils, such as soybean methyl ester (*Glycine max*), in mixtures with conventional jet fuel. Dunn highlighted technical and regulatory challenges, including the need to improve the freezing point of biodiesel to meet aviation specifications. He proposed the "winterization" process, consisting of slowly cooling the biodiesel and subsequently removing the solid crystals formed, reducing the cloud point of the product, which allows it to be mixed with conventional fuel without compromising its essential properties. Despite its technical and scientific relevance, the article had a greater impact on studies on biodiesel production than on research related to the development of alternative fuels for aviation.

From 2010 onwards (point B - Figure 4), there was a significant increase in publications on biokerosene or bio-jet, with eight articles published between 2010 and 2011. This period coincides with the beginning of meetings led by the International Civil Aviation Organization (ICAO), which sought to create a market for offsetting and reducing greenhouse gas emissions in the aviation sector [38]. Between 2010 and 2015, the growth in publications was gradual, but accelerated after 2016 (point C - Figure 4), continuing to grow until 2024 and is expected to continue next year. This increase is possibly related to the implementation of the Carbon Offsetting and Reduction Scheme for International Aviation (CORSIA) [39], a market-based measure (MBM) mechanism created during the 39th Session of the ICAO Assembly. CORSIA has reinforced its commitment to achieving net zero carbon emissions in the aviation sector by 2050.

Figure 4 shows the results related to the number of publications on SAF and production/production technology over the years. There has been a significant increase in recent decades, especially after 2015, which may have been caused by the commitments to reduce greenhouse gas emissions under the Paris Agreement (2015). However, since 2010, CORSIA has also been discussing issues related to the possibility of reducing emissions in aviation. In addition, with the need to enter the carbon market, airlines such as United Airlines, Lufthansa and British Airways have begun to invest in SAF, thus establishing partnerships with biofuel companies and technology startups, which has driven more research and development in the sector.

The graph clearly demonstrates how the scientific community is keeping pace with both market demands and public policy goals, highlighting a strong correlation between academic progress in the field of sustainable aviation fuels and the regulatory and/or economic incentives associated with climate change mitigation in the sector.

## 4. Technological Routes for SAF Production

Due to the increased interest in SAF, several technologies have emerged to meet this demand. The guidance document for biofuel certification created by the International Sustainability & Carbon Certification (ISCC) focuses on wood-based fuels. It mainly deals with the verification of compliance with social, environmental and traceability criteria for biofuels, in accordance with the objectives set by the European regulations for transport fuels. This document compiles sustainability requirements, greenhouse gas (GHG) emissions estimates, and value chain traceability for biofuels [7].

The American Society for Testing and Materials (ASTM) guarantees the technical and safety conditions for the flights. The two main standards are the Standard Specification for Aviation Turbine Fuel (ASTM D1655) and the Standard Specification for Aviation Turbine Fuel Containing Synthesized Hydrocarbons (ASTM D7566), which ensure a fuel's eligibility for use in flights and its traceability element. Recently, ASTM D7566 was updated to include the approval of sustainable aviation fuels (SAF) for passenger and cargo jet engines, with certain restrictions. These fuels must demonstrate good performance and reduced environmental impacts compared to traditional fossil fuels [14], [15], [16], [17], [18]. **Table 1** presents the technological pathways that have already been approved by ASTM.

**Table 1 -** Gasification/Fischer-Tropsch

| Fuel Name | Method | Feedstock | Date certified | TRL | Maximum blend level | Companies Producing SAF | References |
|---|---|---|---|---|---|---|---|
| Fischer-Tropsch – Synthetic paraffinic kerosene (FT-SPK) | Gas-to-liquid conversion process that transforms natural gas or biomass into liquid fuels, divided into two stages: Gasification and Fischer-Tropsch Synthesis. | MSW, agricultural and forest residues | 2009 | 6-8 | 50% | Fulcrum Bioenergy[1]; Red Rock Biofuels[1]; Shell | [5] |
| Fischer-Trosch – Synthetic paraffinic kerosene with added aromatics (FT-SPK/A) | Is a variant of the FT-SPK process that includes the addition of aromatics to the final product | MSW, agricultural and forest residues | 2015 | 6-8 | 50% | | [5] |
| Hydroprocessed Esters Fatty Acids - Synthetic | The esters and fatty acids are hydrogenated and deacidified to form the SAF. | Vegetable and animal fats and oils | 2011 | 8-9 | 50% | Neste; World Energy; SkyNRG; Phillips 66 | [7] |
| Hydroprocessing of Fermented Sugars-Synthetic Iso-Paraffinic fuels (HFS-SIP) | The sugars are fermented into alcohols, undergo hydrogenation, and are converted into iso-paraffinic hydrocarbons. | Sugars | 2014 | 7-8 | 10% | | [5] |
| Alcohol-to-Jet Synthetic Paraffinic Kerosene (ATJ-SPK) | After fermentation produces alcohol, dehydration and polymerization occur to form larger hydrocarbons, followed by hydrogenation. | Starches, sugars, cellulosic, biomass | 2016 | 7-8 | 50% | Gevo Inc.; Lanzajet | [5] |
| Co-processing of up to 5 vol% fats and oil in a refinery to produced kerosene | | Vegetable and animal fats and oils | 2018 | - | 5% | | [7] |
| Catalytic Hydrothermolysis Synthetic Kerosene (CH-SK) | Biomass is subjected to high temperatures and pressures in the presence of a catalyst to form a mixture of oils and organic compounds, followed by dehydration and hydrogenation. | Renewable fats oils and greases | 2020 | 6 | 50% | | [7] |
| Hydroprocessed Esters and Fatty Acids Synthetic Paraffinic Kerosene (HC-HEFA-SPK) | | Hydrocarbon-rich algae oil | 2020 | 6 | 10% | | [5] |

Analysis of the various routes and raw materials used in SAF production reveals significant adaptability to local resources and regulations. In the context of the European Union, decision-making is influenced by three main factors: the legal framework of government incentives and $CO_2$ taxes, the availability of raw materials, and the maturity of SAF production routes. This approach is in line with global decarbonization and sustainability goals, promoting a cleaner and more sustainable future for the aviation industry. In addition, collaboration between companies, governments, and research institutions will be crucial to address the technical and economic challenges that still exist in the large-scale implementation of SAF.

## 5. Current Mandates and Policies

In the context of commercial aviation, the mandates currently in force, especially in the European Union, together with the incentives implemented in the United States, the United Kingdom and other countries, establish the gradual introduction of renewable biofuels, known as SAF, in the aviation fuel pool. These mandates vary, but in general call for an initial introduction of 2% SAF from 2025, with a gradual increase to between 6% and 10% by 2030, as well as an ambitious goal of carbon "neutrality" by 2050. However, the production volumes currently available are far below the amount needed to meet these mandates. Ongoing initiatives related to SAF production around the world are still far from producing the volumes required to achieve the established objectives. Therefore, it is critical to understand the technological, regulatory, and economic constraints that limit SAF production in order to identify the bottlenecks and obstacles that need to be overcome in order to effectively comply with these mandates [19].

**Table 2** illustrates the different approaches adopted by Brazil, the European Union and the United States for the use of SAF in aviation, highlighting the specific regional strategies to achieve decarbonization goals and encourage the use of sustainable fuels in the sector. Each region has policies that combine financial incentives, regulations and emission reduction targets, reflecting a commitment to promoting more sustainable aviation in different regulatory and economic contexts.

**Table 2 - Approaches to achieve decarbonization goals and encourage the use of sustainable fuels in aviation.**

| Criterion | European Union | United States | Brazil |
|---|---|---|---|
| **Main Regulation** | ReFuelEU Aviation Regulation (parte do Fit for 55) | No mandatory federal mandates for SAF; tax and state incentives (e.g. California LCFS) | Fuel of the Future Program, ProBioQAV (Law No. 14,993/2024) |
| **SAF Blending Targets** | 2% in 2025, with a target of 70% in 2050 | Varied incentives, without specific mandates | Gradual reduction of GHG emissions, starting with 1% in 2027 and reaching 10% in 2037 |
| **Synthetic Fuels** | 0.7% of synthetic fuels in 2030, with a target of 35% in 2050 | Synthetic SAF incentivized by tax credits, but no defined targets | |
| **Sustainability Criteria** | Renewable Energy Directive (RED) compliance | SAF must comply with ASTM D7566 standard for safety and mixing | Life cycle analysis (LCA) of the fuel; regulation by the ANP |
| **SAF Infrastructure** | Infrastructure for SAF storage and mixing mandatory at airports | Federal and state infrastructure grants, including the FAA's FAST program | Incentive to SAF logistics and distribution |
| **Key Incentives** | Mandatory mixing; Staggered goals | SAF tax credit ($1.25 to $1.75/gallon) via the Inflation Reduction Act (IRA); LCFS credits in California | Staggered GHG reduction targets, incentives for $CO_2$ capture and geological storage |
| **Other Initiatives** | | Sustainable Skies Act: proposta de crédito fiscal adicional de até $1,50 por galão para SAF | |
| **Existing Program Integration** | Fit for 55 | California LCFS and State Tax Credits | RenovaBio, Mover Program, Proconve |
| **Final Emissions Goal** | Carbon neutrality by 2050 | Incentives to achieve emission reductions; No federal carbon neutrality mandate | Gradual reduction of emissions, without a neutrality mandate |

In Australia and New Zealand, a target of reducing aviation fuel emissions by 2.5% by 2025, 3% by 2030 and 10% by 2050 was proposed in 2022. These targets are less ambitious compared to European guidelines, which set higher blending quotas to accelerate aviation decarbonization [20], [21].

The Brazilian scenario for the use of SAF has been structured by the federal government since 2021, when the Fuel of the Future Program (CNPE Resolution No. 7/2021) was created, with an inter-ministerial technical group to evaluate the insertion of SAF in the transport matrix. In 2023, the program resulted in a bill that proposes the creation of the National Sustainable Aviation Fuel Program – ProBioQAV. This program establishes gradual emission reduction targets to be achieved with the use of SAF, starting at 1% in 2027 and progressively increasing to 10% in 2037, with life cycle analyses of production routes. Both ProBioQAV and CORSIA (Carbon Offsetting and Reduction Scheme for International Aviation) aim to reduce emissions, but are based on carbon reduction targets rather than volumetric mandates. While ProBioQAV applies to domestic flights, CORSIA defines a carbon-neutral growth by 2035, with the goal of net-zero emissions in international aviation by 2050. Both programs encourage the use of SAF as an essential tool to achieve their environmental goals [22], [23].

Among the studies carried out on the economic, political and technological aspects of production, it was found China is the country with the largest number of published works as can be shown in **Table 3**. This factor can be explained by the policy applied since 2015, after the Paris Agreement, where the country committed to reducing its emissions. In addition, as of 2020 and 2021, it presented “double carbon” and “14th Five-Year Plan” goals that reinforce these guidelines, with the prospect of achieving $CO_2$ emissions neutrality by 2060. One of the goals is to implement a 4.5% reduction in $CO_2$ emissions per ton-kilometer of air traffic, compared to 2020. In addition, China has carried out test flights at Hongqiao Airport in Shanghai since 2013, and SAF began to be used on commercial flights in 2015. By 2024, Sinopec, one of the largest oil companies in the country, registered the technology for the production of biofuels for transport and the country began testing the use of SAF in the largest airlines, such as Air China and China Southern Airlines, until 2025 [24], [25], [26].

**Table 3 - Country Frequency Distribution for SAF-Related Keywords**

| Key-words for search | ("bio kerosene" OR "biojet" OR "SAF") AND "economic" NOT ("catalytic" OR "oxidation" OR "production") | | ("bio kerosene"OR "biojet" OR "SAF") AND "production" NOT ("economic") | | ("bio kerosene" OR "biojet" OR "SAF") AND "production" AND ("HEFA" OR "Alcohol-to-Jet" OR "Fischer-Tropsch") | |
|---|---|---|---|---|---|---|
| | **Country** | **Freq** | **Country** | **Freq** | **Country** | **Freq** |
| | USA | 186 | CHINA | 4707 | USA | 186 |
| | CHINA | 113 | USA | 1902 | CHINA | 113 |
| | BRAZIL | 74 | SPAIN | 1129 | BRAZIL | 74 |
| | GERMANY | 54 | INDIA | 688 | GERMANY | 54 |
| | UK | 36 | BRAZIL | 608 | UK | 36 |
| | NETHERLANDS | 35 | ITALY | 579 | NETHERLANDS | 35 |
| | ITALY | 22 | UK | 474 | ITALY | 22 |
| | SWITZERLAND | 21 | IRAN | 461 | SWITZERLAND | 21 |
| | CANADA | 19 | SOUTH KOREA | 415 | CANADA | 19 |
| | MEXICO | 18 | FRANCE | 385 | MEXICO | 18 |

## 6. Demand Estimation to reach PROBIOQAV and CORSIA goals:

The **table 4** below summarizes a possible combination of the three scenarios described, aiming to meet both ProBioQAV and CORSIA goals with CAPEX estimations based on announced projects and other public data:

**Table 4 - Possibilities of scenarios to meet the ProBioQAV and CORSIA targets**

| Scenario | Description | SAF Production Capacity | Feedstock | Emissions Reduction Target | Future Investment |
|---|---|---|---|---|---|
| **Scenario I** | Committed projects from companies such as BBF, Acelen and Petrobras | BBF: 250 thousand m³/year (2026) Acelen: 500 thousand m³/year (2027) Petrobras: 350 thousand m³/year (2029) | Oil of palm, soy, milho, bovine tallow, macaúba (future) | Achieves, on average, 38% of CORSIA and ProBioQAV targets | R$ 8,7 billion |
| **Scenario II** | Explores the use of traditional and alternative raw materials | From 3.7 to 8 million m³/year by 2037 | Soybean oil, sugarcane and corn ethanol, cellulosic ethanol, macaúba oil, agave ethanol | Accounting for 36% to 78% of aviation fuel demand by 2037 | R$ 21 e R$ 48 billion**.** |
| **Scenario III** | Uses organic waste from agro-industrial activities | Undefined potential, depending on the full use of available waste | Tallow bovino, resíduos de cana, resíduos de eucalipto | It could meet up to 82% of emissions reduction targets by 2037 | R$ 13,6 billion. |

Considering the goals established by PROBIOQAV and CORSIA, the EPE has published in 2024 and estimation of SAF demand considering different Carbon Offset Indexes (CI) for SAF production, as per **table 5** below:

**Table 5 - Expected demand for SAF over the years according to EPE (2024)**

| **Year** | **CORSIA Goal** | | **ProBioQAV goal** | | **Total Demand** | |
|---|---|---|---|---|---|---|
| | **Lower CI** | **Higher CI** | **Lower CI** | **Higher CI** | **Lower CI** | **Higher CI** |
| | kt/y | kt/y | kt/y | kt/y | kt/y | kt/y |
| 2027 | 721.6 | 2307.2 | 32 | 100.8 | 753.6 | 2408 |
| 2029 | 949.6 | 2992 | 112.8 | 356.8 | 1062.4 | 3348.8 |
| 2034 | 1347.2 | 4244 | 292.8 | 921.6 | 1640 | 5165.6 |
| 2037 | 1864.8 | 5874.4 | 444.8 | 1400 | 2309.6 | 7274.4 |

**Source:** EPE Sustainable aviation fuel 2024 and authors analysis

Besides the uncertainties of the estimations, these figures suggest the order of magnitude of investment necessary to reach the SAF goals defined in the local legislation (ProBioQAV) and in the international goals established by CORSIA

## 7. Economic Evaluation of HEFA-SPK and ETJ routes

Considering the level of technological readiness and the commercial availability of feedstock, a high-level analysis comparing the variable cash cost of SoyBeans based HEFA route and Ethanol 1G based ETJ (Ethanol to Jet) route is shown below, looking their relative competitiveness based on historical prices of feedstock, energy and products in the Brazilian market. For simplification, by products of both routes were assumed to be sold for nafta price. SAF price was assumed to be the same as conventional jet fuel price. Ethanol, Soybeans oil and other inputs assumed as per market prices. In summary, neither subsidies nor incentives were considered in the analysis. Catalyst costs and other variable costs other than feedstock and main energetic costs were not considered due to the relatively low relevance of it compared to the other elements in the analysis. Technical data for the analysis were based on the references [27] and [28], that simulate Alcohol-to-Jet and HEFA routes for SAF production in the Brazilian context (Table 6).

**Table 6 - Elements for analysis comparing the variable cash cost of the soybean-based HEFA route and the 1G ethanol-based ETJ (ethanol to jet) route**

| Elements analyzed | Unit of measurement | HEFA-SPK | ETJ |
|---|---|---|---|
| Soybeans oil | Kg | 2,0200 | - |
| Ethanol Consumed | Kg | - | 3,2411 |
| Hydrogen Consumed | Kg | 0,0800 | 0,0398 |
| Electricity Consumed | KW/h | 0,7100 | 0,2377 |
| Heat consumed as natural gas | Kg | 0,8926 | 0,1897 |
| SAF produced | Kg | 1,0000 | 1,000 |
| By products Produced considered as Naphta | Kg | 0,8200 | 0,8961 |

Figure 5: [A] Total Variable Cost Product with product; and [B] Exchange rate.

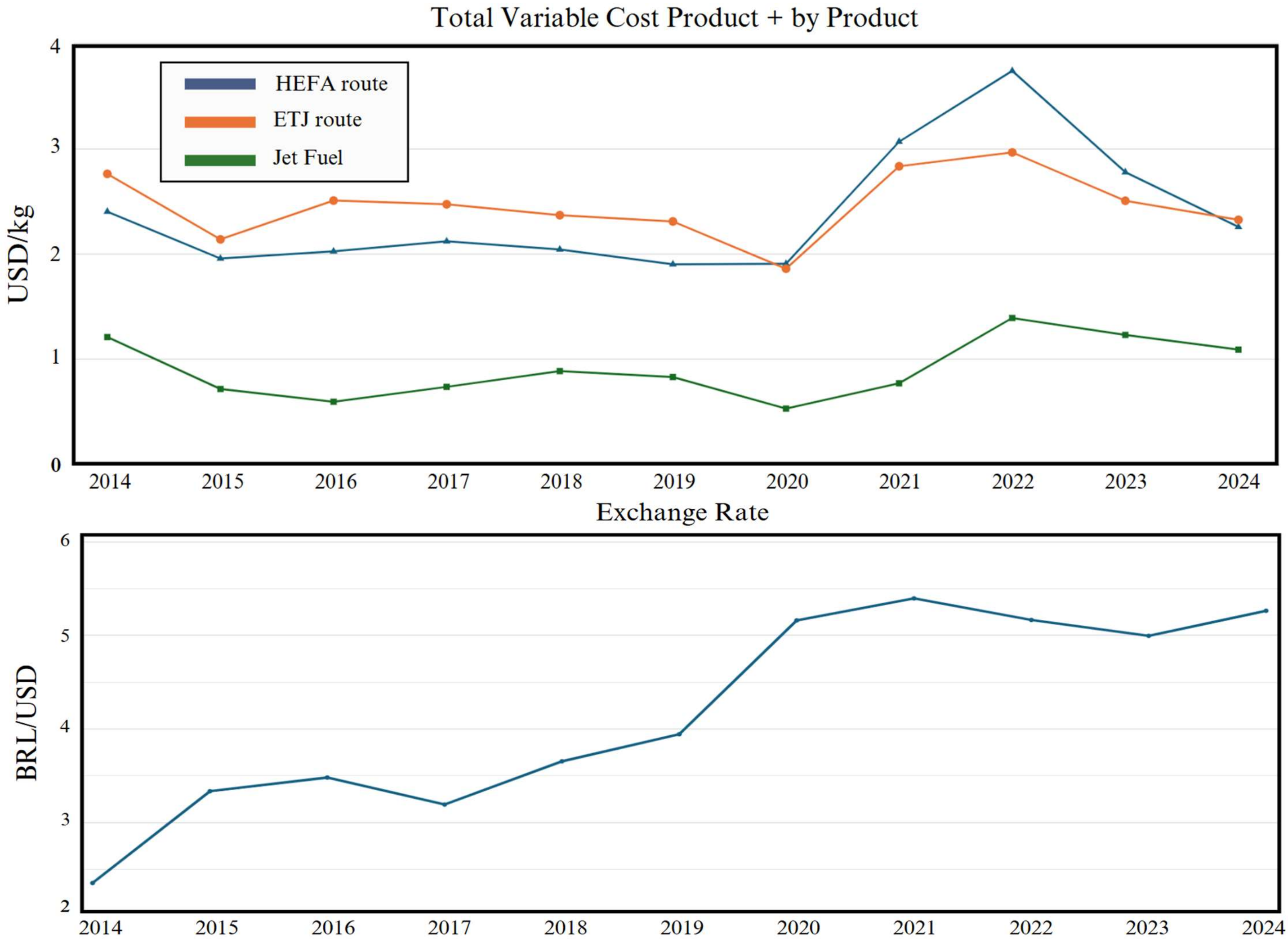

Figure 6 - Variable Cost comparison HEFA x ETJ routes and Jet Fuel Price

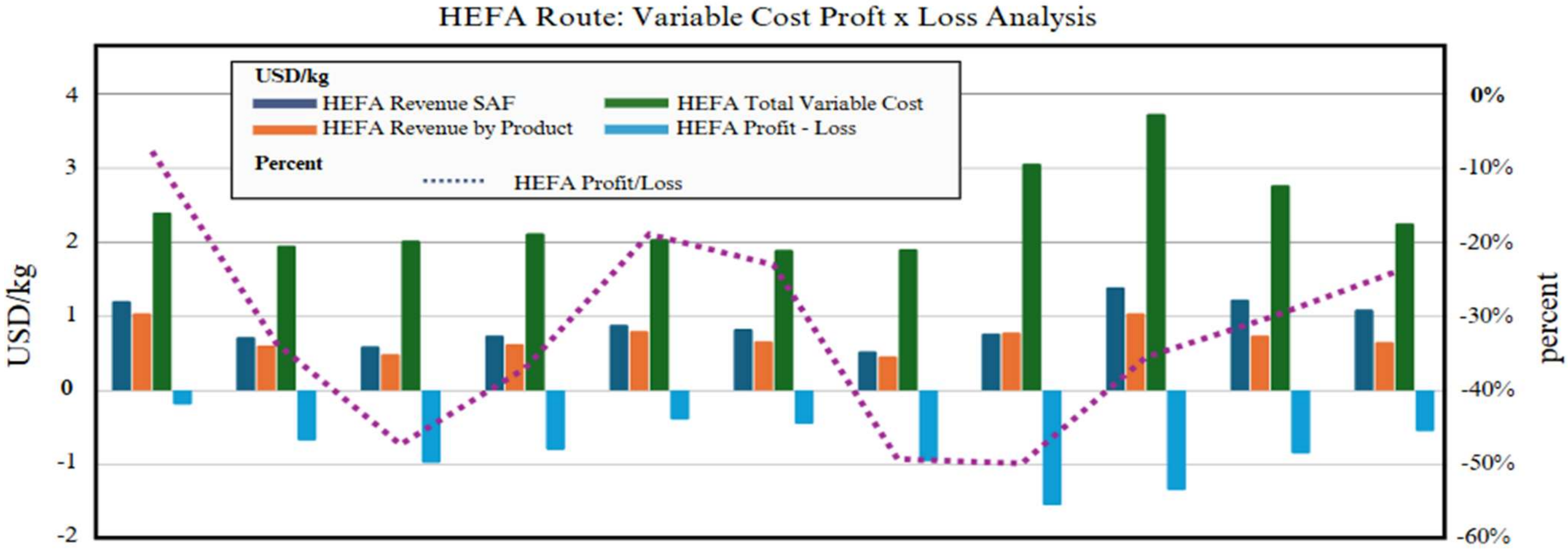

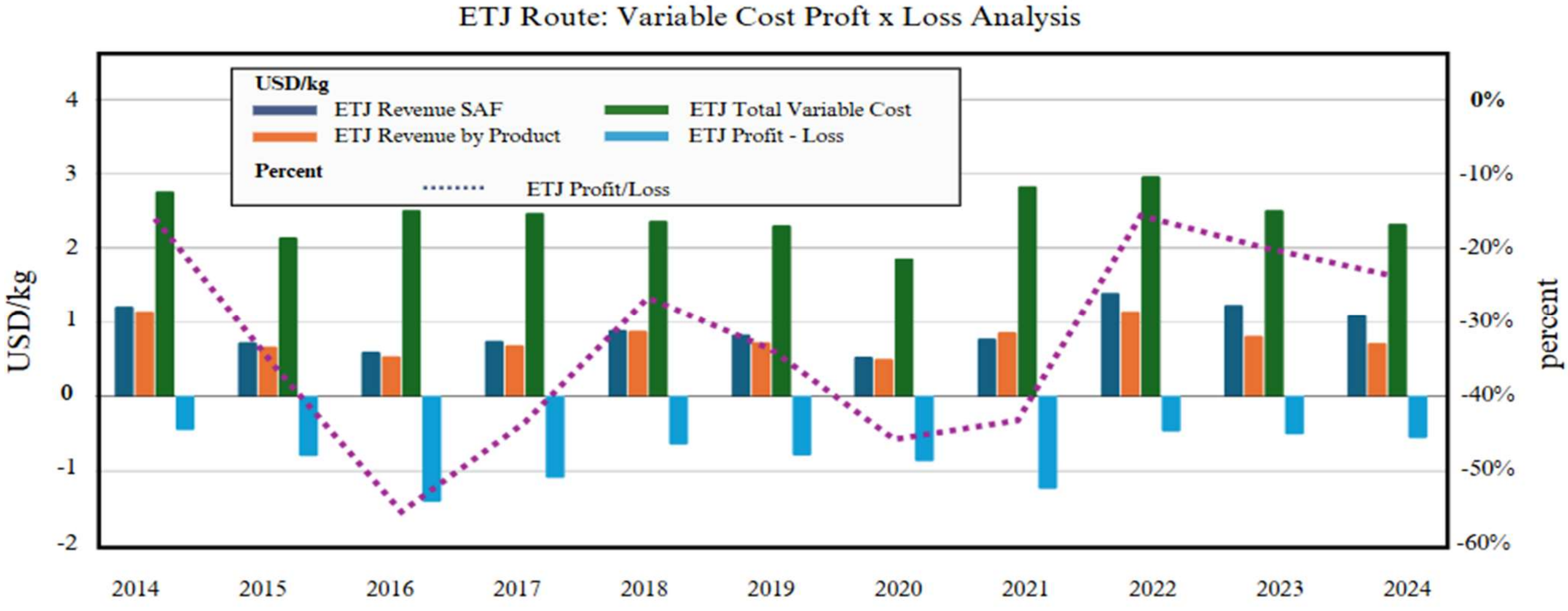

The market prices for the 2014-2024 timeframe are shown in **Table 7** and the figures below summarize the comparative results. The results below indicate that both routes would generate SAF at variable cost above historical jet fuel market prices.

As expected, production costs are driven mostly by feedstock, but there is no clear indication that the cost drivers are the same for both routes in the timeframe of this analysis. From an investors perspective, economic incentives would be required for an investment in either route as discussed in the session Incentive Policies below.

**Table 7 -** Price of main feedstock and products and by products

| Year | Ethanol Price R$/kg | Soybeans Oil Price R$/kg | Jet Fuel Price R$/kg | Nafta Price R$/kg | H2 Price R$/kg | Natural Gas R$/kg | Electricity R$/kWh |
|---|---|---|---|---|---|---|---|
| 2014 | 1.57 | 2.04 | 2.42 | 2.55 | 5.84 | 0.41 | 0.39 |
| 2015 | 1.70 | 2.30 | 2.04 | 2.10 | 5.84 | 0.32 | 0.42 |
| 2016 | 2.11 | 2.49 | 1.77 | 1.77 | 5.84 | 0.42 | 0.40 |
| 2017 | 1.91 | 2.45 | 2.00 | 2.05 | 5.84 | 0.38 | 0.35 |
| 2018 | 2.08 | 2.65 | 2.76 | 3.02 | 5.84 | 0.46 | 0.40 |
| 2019 | 2.19 | 2.63 | 2.78 | 2.72 | 5.84 | 0.40 | 0.38 |
| 2020 | 2.26 | 3.54 | 2.33 | 2.44 | 5.84 | 0.34 | 0.36 |
| 2021 | 3.71 | 6.59 | 3.58 | 4.36 | 5.84 | 0.69 | 0.41 |
| 2022 | 3.70 | 7.82 | 6.10 | 5.60 | 5.84 | 1.06 | 0.44 |
| 2023 | 3.01 | 5.38 | 5.22 | 3.83 | 5.84 | 0.63 | 0.46 |
| 2024 | 2.92 | 5.05 | 4.88 | 3.56 | 6.40 | 0.63 | 0.48 |

The literature also supports the conclusion of this evaluation, and reinforces the need to analyze each potential SAF project in light of market conditions especially in regards to feedstock price and availability, electricity price, carbon reduction policies and logistics. In example, according to Braun, Grimme and Oesingmann [20], despite the regulatory efforts, the widespread adoption of SAF will depend on competitiveness in the marketplace against other $CO_2$ reduction alternatives such as cap-and-trade, carbon credits and book-and-claim schemes. The study looked at various conversion pathways, raw materials, or input cost scenarios. In total, 230 comments on minimum sales prices (MSP´s) and costs of production were included in the analysis. According to the study, the variation in production costs can be attributed to several factors such as the efficiency of processes (e. g. conversion yields), feedstock and electricity prices among others.

For PtL fuels, electricity prices determine the minimum selling prices to a large extent. Assumptions on electricity prices themselves vary across different studies, therefore also impacting the model results in the respective studies. Moreover, the provision of sufficient electricity from renewable sources would require substantial capital expenditures since it needs to be additional, in order not to compete with alternative uses required in the transformation of power systems. It also favors locations where an efficient combination of different renewable power sources such as photovoltaics, hydro and wind are available.

Other studies such as Wang et al [29], make similar analysis for the relative competitiveness of the SAF pathways and besides assumptions and base case scenarios are not the same, the general conclusion is that MSP for SAF is most likely to be way above conventional jet fuel market price meaning that there is no economic return in this industry without policy incentives. In addition, evaluates the impact of some example of incentive policies such as: feedstock subsidies, capital grants, output based incentives, and also policies intended to reduce project risk for six SAF pathways: isobutanol to jet (ATJ) from corn grain, hydroprocessed esters and fatty acids (HEFA) from inedible fats and oils (IFO), HEFA from palm fatty acid distillate (PFAD), synthesized iso-paraffins (SIP) from sugarcane, Fischer-Tropsch (FT) gasification and synthesis from municipal solid waste (MSW), and micro FT from wood residues are modeled as individual refineries using harmonized financial assumptions such as capital investment, capital cost, depreciation, taxes, inflation among other standard project financial evaluation. However, the differences among the pathways under evaluation such as production throughput, Capital Investment per liter produced, by products production and feedstock availability need to be considered when evaluating the study conclusions.

## 8. Assessment of Potential Incentive Policies

Considering the challenges to reach economic sustainability to the SAF industry, the literature explores potential incentives and levers that could increase its attractiveness. One example can be found in the study [29], based on biofuel policies implemented in different jurisdictions around the world. This article evaluates four different incentive policy types:

**Output Based Incentives:** An output based incentive is a policy for which the fuel producer receives some monetary benefit tied to the quantity or type of fuel produced and sold. For example, the value of the benefit could be a function of production volumes, which is the case for Renewable Identification Numbers (RINs) generated under the US RFS2. It could also depend on the life cycle emissions reductions compared to petroleum fuel, which is the case for the California LCFS and ICAO CORSIA policies.

**Feedstock Subsidies:** A feedstock subsidy is a monetary benefit to reduce a facility's operating costs for feedstock. It may also have the benefit of supporting feedstock producers, by providing agricultural incentives to establish the supply chain. A feedstock subsidy could also take the form of a monetary credit or avoided cost for using waste products such as MSW that would otherwise take up landfill capacity. In the case analyzed in this article, taxes over feedstock and other inputs were used to simulate this impact on the cost of SAF.

**Capital Grants:** A capital grant is typically a one-time monetary benefit, granted by the government to cover or reduce facility construction costs. Two examples of implemented capital grant policies are the 2018 Calrecylce Organics recycling projects and the Rural Energy for America Program (REAP).

**Risk Reduction Policies:** Two additional policies, loan guarantees and offtake agreements, are modeled to quantify the impact of risk reduction policies on project economics. A loan guarantee is an agreement between the guarantor and the bank that states if a refinery defaults on a loan, the guarantor will pay the bank in its stead. This results in a lower cost of debt, as some risk associated with the project has been borne by the guarantor (usually the government), rather than the bank. REAP also provides loan guarantees up to $25 million. A loan guarantee

is modeled here as a reduction in the cost of debt. An offtake agreement occurs when a fuel purchaser agrees to purchase fuel quantities at a pre-negotiated price at some future date rather than the prevailing market price. A number of airlines have established offtake agreements with SAF producers, such as Lufthansa and Gevo, and United Airlines and Fulcrum bioenergy. In the present article, this was modeled using a premium over the market price of SAF and byproducts.

Brazilian Context: As nowadays the Brazilian tax framework has a significant impact on the SAF cost structure, this work has evaluated options for the implementation of the above-mentioned potential incentives, in the Brazilian context. One possible option would be the use of alternative taxation systems already used in different industries facing similar challenges. One option evaluated is the concept of Regulatory Sandbox, explained below:

According to the Gonçalves [30], Sandbox is the term used by technology and information (IT) professionals to define a controlled, isolated, and safe testing environment from external interventions, without interrupting other existing applications. In view of the strong growth of fintechs and the regulatory challenges it has raised in relation to "Prudential and risk-based regulation", the Financial Conduct Authority (FCA) has developed in 2015 the first model of the sandbox aiming to foster the innovation of products, services and business models through digital-technology. In light of the good results obtained by this first model, other institutions and sectors started to try the same model. In example, a sandbox in the financial, capital and insurance segments was created in 2019 by the Brazilian Ministry of Economy, the Central Bank (BC), the Securities and Exchange Commission (CVM) and the Superintendence of Private Insurance (Susep). Another example, in 2020, the Brazilian CVM enacted CVM Instruction No. 626, which was revoked by CVM Resolution No. 29/2021, aiming to regulate the constitution and operation of the regulatory sandbox for legal entities in the securities market. The purpose of the sandbox is to serve as an instrument to encourage innovation in the capital market, guide participants on regulatory issues during the development of the activities in order to increase legal certainty, reduce costs and time in the product and service development, increase competition between service providers, in addition to improving the "framework applicable to regulated activities".

This "regulated self-regulation", originated from the difficulties of the traditional state standardization of economic activity in the face of risks that new technologies, enables the communication between the various agents of society (state, private, non profit organizations, etc.) in order to create strategies for governance for the application of more effective rules.

According to the Brazilian Federal Court of Accounts (TCU), the environment experimental experiment generated with regulatory sandboxes aims to suspend, for a determined period, "the obligation to compliance with standards required to operate in certain sectors, allowing companies to enjoy a differentiated regime to launch new innovative products and services in the market, with less bureaucracy and more flexibility, but with the monitoring and guidance from regulatory bodies" . According to this assumption, it would be possible for a regulatory body granting temporary authorizations, suspend for a predefined period the applicable legislation, making it flexible, in order to enable the performance of strictly pre-selected entities and/or companies, for a restricted number of clients, in a whose main purpose is the development of new business models based on the use of existing or innovative technologies, within the limits of the legislation. The initiatives for the implementation of the regulatory sandbox were intensified with the Law Complementary No. 182/2021 (Legal Framework for Startups), which introduced in Brazil, in general, the legal institute of the sandbox allowing the use of the tool for the production of rules and regulations in experimental environments beyond the financial market in the country.

Considering the negative economic results shown previously for both HEFA and ATJ routes in the Brazilian market context, the authors of this work have interviewed several stakeholders in the industry such as potential producers, potential customers and policy makers, aiming to identify levers and incentives that, in the perspective of these stakeholders, could enhance SAF economic attractiveness and incentivize investments in this area. The highlights of these interviews were: The need of subsidized or granted capital for investment in the pioneer plants at least, the need of a special tax environment for the SAF value chain at least for a determined period, the need of regulation and predictiveness for the carbon credits market and the need for premium prices for SAF and by products in comparison with conventional products. Besides based on a limited number of interviews, the findings are corroborated by the literature, as mentioned before in this article.

Based on this assessment, simulations were carried out in order to evaluate the impact of the following potential incentive policies to the economic viability of SAF:

In the present study, simulations were carried out for HEFA and ATJ routes, using Brazilian market conditions for feedstock and products, assuming the applicability of the regulatory sandbox concept for the SAF industry in order to evaluate the impact of some potential incentives and market levers to economic viability of SAF:

**1) Tax Reduction Impact:** Assuming the applicability of the regulatory sandbox concept to the SAF industry, the effect of tax reduction in the feedstock and energy supplies required for SAF production were evaluated.

The **Table 8** below shows the taxes charged for the different SAF production inputs and products in the state of São Paulo in 2024 as an example, since taxes may vary from one state to another in Brazil.

**Table 8 -Taxes on different inputs and products of SAF in the state of São Paulo - Brazil in 2024**

| Product | ICMS (São Paulo) | PIS/PASEP (Federal) | COFINS (Federal) | CIDE (Federal) |
|---|---|---|---|---|
| Soybean Oil | 12% | 1,65% | 7,6% | Not applicable |
| Ethanol | 18% | 1,86% | 8,6% | Not applicable |
| Natural Gas | 18% | 1,86% | 8,6% | Not applicable |
| Electricity | 18% | 1,65% | 7,6% | Not applicable |
| Aviation Kerosene | 18% | 1,65% | 7,6% | Not applicable |
| Nafta | 18% | 1,65% | 7,6% | Not applicable |
| Diesel | 18% | 1,86% | 8,6% | R$ 0,10/L |

For simplification the simulations assume a case of 50% tax reduction and another one at 100% reduction for all inputs.

**2) Carbon Price:** Brazil is a country with a large natural capacity to generate environmental assets, mainly carbon credits, subject to national and international transactions. There are several legal rules in Brazil that provide for the existence of carbon assets. At the Federal level, the most relevant standards are: National Policy on Climate Change (PNMC), Law No. 12,187, of December 29, 2009, the National Policy for Biofuels (RENOVABIO), law 13576/2017 and the Forest Code, law 12.651/12.

The law RENOVABIO defines relevant elements for the Brazilian carbon market, among them: a) **Biofuels Certification:** a set of procedures and criteria in a process, in which the inspecting firm evaluates the conformity of the measurement of aspects related to the production or import of biofuels, as a function of energy efficiency and greenhouse gas emissions, based on a life cycle assessment; b) **Life cycle:** consecutive and chained stages of a product system, from the raw material or its generation from natural resources to the final disposal, as defined in regulations; c) **Decarbonization Credit (CBIO):** instrument registered in book-entry form, for the purpose of proving the individual target of the fuel distributor. Such credits can be traded in organized markets, including auctions.

However, to date, there is no regulation of the Brazilian Market for Emissions Reduction (Mercado Brasileiro de Redução de Emissões - MBRE) provided for in Law No. 12,187, of December 29, 2009, as well as we have not created a legal framework appropriate to dealing with and encouraging transactions with carbon assets. The Bill 20258/2021 (Bill of Carbon), under analysis by the senate in November 2024, aims to present the regulation of the Brazilian Emissions Reduction Market (MBRE), determined by the National Climate Change Policy – Law No. 12,187, of December 29, 2009, with a view to, among others, to: **a)** Conceptualizing and determining the legal nature of carbon assets (carbon credits); **b)** Establish a system for recording the inventory of greenhouse gas emissions and the national accounting of emission reductions and their transactions; **c)** Determine the fungibility of carbon assets, in order to establish the interoperability of different market mechanisms on greenhouse gas emission reductions, with the appropriate technical-scientific standardization; **d)** Establish the domestic market for emission reduction, based on our NDC, the national inventory and the characteristics of our economic sectors; **e)** The promotion of the activities of projects for the reduction and removal of greenhouse gas emissions.

For the purpose of this article, it is assumed that the future legal framework will allow the fungibility of different carbon assets. So, the effect of $CO_2$ price is simulated as a credit in the revenue of SAF production using the CBIO price as reference. In order to calculate this impact, a carbon index of 89 $gCO_2/MJ$ is assumed for conventional jet fuel, a carbon index of 36 $gCO_2/MJ$ is assumed for ATJ based SAF and 42,9 $gCO_2/MJ$ is considered for HEFA based SAF, using Core LCA assumptions, without Land Use Change (LUC) Impact and simplifications as per [31]. In this study the indicator used was the price of the CBIO´s , created by RENOVABIO and currently traded in the Brazilian fuel market. The CBIO is considered as equivalent to one tCO2.According to trade data available at [32], the average price of CBIO in 2024 (until October) was 89 R$/CBIO. For the sake of this study sensitivity analysis, the CBIO price is assumed at 100 R$/tCO2, and a second scenario is calculated assuming 400 R$/tCO2. CO2 emissions in feedstock, Land Use Change (LUC) and/or SAF transportation are not considered in this study.

**3) Risk Reduction - Premium Price for SAF:** Considering the mandates in place, the air companies will be required to mitigate their GHG emissions and it is considered reasonable by the stakeholders interviewed during the elaboration of this study that commercial agreements between producers and consumers of SAF could use a premium over conventional jet fuel price in long term supply contracts. For the sake of exercise, premiums of 25% and 50% over SAF market price were considered in the simulations.

**4) Risk reduction - Premium Price for By Products:** Considering the relevant amount of by products produced by both routes, and assuming that the chemical and petrochemical industry worldwide is keen to acquire low carbon and biobased products, the by product streams were valued using fossil nafta price as a reference and the same premium range used for SAF was applied in the simulation. Similar commercial framework described in item 3 is considered reasonable by the stakeholders interviewed for this study.

**5) Capital grants:** In order to evaluate the order of magnitude of capital grants necessary to generate neutral NPV for SAF projects, a reverse discounted cash flow analysis was performed for the scenarios described below. Capital grant mechanisms are frequently funded by the Brazilian Development Bank (BNDES) and one example of a call for projects is found in [33].

In all cases, the the reference prices considered were the average market price in the period of 2021 to 2024 as per **table** below:

**Table 9 - Average prices between 2021-2024 for subsidies and final energy product**

| Year | Ethanol Price R$/kg | Soybeans Oil Price R$/kg | Jet Fuel Price R$/kg | Nafta Price R$/kg | H2 Price R$/kg | Natural Gas R$/kg | Electricity R$/kWh |
|---|---|---|---|---|---|---|---|
| 2021 | 4.37 | 6.59 | 4.17 | 5.17 | 6.90 | 0.82 | 0.49 |
| 2022 | 4.37 | 7.82 | 7.20 | 6.59 | 6.90 | 1.25 | 0.52 |
| 2023 | 3.55 | 5.38 | 6.16 | 4.50 | 6.90 | 0.75 | 0.54 |
| 2024 | 3.44 | 5.05 | 5.75 | 4.20 | 7.55 | 0.75 | 0.57 |

As some prices include different taxes, those were isolated in the economic analysis. The isolated impact of each of the four incentives analyzed was not enough to generate positive margins in any of the SAF production routes evaluated, so two scenarios were created, to be compared with a base case scenario, without any incentive for SAF value chain:

**Scenario 1)** This scenario evaluates the impact of 50% tax reduction on inputs, a market price of 200 R$/t $CO_2$ eq and a 25% premium for SAF and renewable by-product produced for each route.

**Scenario 1)** This scenario evaluates the impact of 100% tax reduction on inputs, a market price of 400 R$/t $CO_2$ eq and a 50% premium for SAF and renewable by product for each route.

The Figure 7show the result of the analysis. In this analysis, “margins” are calculated by selling price minus total cost of feedstock and other inputs such as energy and utilities. This concept is known as “contribution margin”. The “net margin baseline” bar indicates the contribution margin generated by SAF and by-products considering market price and taxes as per current situation. The other bars show the impact of each incentive described above in the contribution margin.

**Figure 7. Result for sensitivity analysis for the two production routes (HEFA and ETJ)**

The sensitiveness analysis shows that the major impacts may come from tax reductions on feedstock and other inputs, along with premium prices for SAF and by-products. At the price levels considered, $CO_2$ premium has a positive but less relevant impact. Moreover, the amount and value of the by-products generated in the conversion process may have a significant impact on the economic results of SAF production. Deeper analysis on the actual value of by biogenic products such as naphtha, NGL and diesel, accounted as “nafta” in this study, along with alternatives to control the selectiveness of the production process aiming to maximize the yield of the most desirable products would be relevant elements for further studies.

In order to assess the attractiveness of an investment for SAF production a simplified discounted cash flow (DCF) analysis was done, assuming data available in the references [27], [28], [34]. The main elements of cost in a DCF analysis are CAPEX for the plant construction, OPEX, which is mostly the cost of feedstock and other supplies multiplied by the specific consumption in the process, fixed cost such as labor and maintenance, the opportunity cost of capital for the investor, usually named as weighted average cost of capital (WACC) and the taxes. The revenue streams are the sales generated by the product stream, in this case, SAF and by-products.

The main results are the net present value of the investment (NPV), calculated by the net cash generation discounted by the WACC during a given period and the internal rate of return of the investment (IRR) which is the proxy of WACC that leads NPV to zero.

As majority of the stakeholders interviewed considered the capital costs published in the literature lower than the actual investment cost in similar process plants in the chemical and petrochemical industries and the values published in the announced projects uncertain, a reverse

DCF exercise was made, aiming to identify maximum CAPEX affordable for each of the routes in scenarios 1 and 2, along with the base case (no incentives) scenario. The values of CAPEX presented in the **Table 10**, when positive, indicate the maximum affordable CAPEX for the project to reach NPV = 0 at a WACC of 12%. The negative values indicate the NPV of discounted revenue cash flows even at CAPEX = 0, meaning that for those scenarios, on the top of the incentives mentioned above and considered in scenarios 1 and 2, a capital grant would be necessary to reach NPV = 0. The common assumptions in the reverse DCF analysis are: SAF production capacity: 300 kt/year, WACC: 12%, Plant Useful life for amortization: 20 years, no residual value and no perpetuity were considered in the economic assessment.

**Table 10 - CAPEX values for NPV=0 for the scenarios analyzed**

| | **Maximum CAPEX for NPV = 0** | |
|---|---|---|
| | **HEFA** | **ATJ** |
| **WACC*** | 12% | 12% |
| **Base case*** | -$1.838.641.593 | -$184.292.690,7 |
| **Scenario 1*** | -$307.760.803 | $103.508.811,2 |
| **Scenario 2*** | $1.009.664.966 | $366.277.521,3 |
| **CAPEX Reference (27)*** | $108.751.052 | $20.235.662,46 |
| **CAPEX Reference (34)*** | $ 109.588.512,6 | $109.588.512,6 |

*The average dollar rate for the period 2021 to 2024,. Rate: 1 US Dollar/USD = 5,20 Real/BRL

The **table 11** below summarizes the other elements used in the DCF analysis, for the base case, along with the assumptions for each incentive used in the scenarios 1 and 2.

**Table 11 - Elements for DCF Analysis**

| | Average 2021-2024 | HEFA Route | | ATJ Route | | |
|---|---|---|---|---|---|---|
| | Price US$/un | Un/kg SAF | US$/kg | | Price US$/un | un/kg SAF |
| FOREX avg 2021-2024 R$/US$ | 5.20 | | | | | |
| Variable Cost feedstock | 1.08 | 2.02 | 2.19 | 0.68 | 3.24 | 2.19 |
| Tax on feedstock | 0.23 | 2.02 | 0.46 | 0.19 | 3.24 | 0.62 |
| Variable cost electricity | 0.09 | 0.71 | 0.07 | 0.09 | 0.24 | 0.02 |
| Tax on Electricity | 0.03 | 0.71 | 0.02 | 0.03 | 0.24 | 0.01 |
| Variable Cost Hydrogen | 0.10 | 0.08 | 0.01 | 0.10 | 0.04 | 0.00 |
| Tax on Hydrogen | 0.03 | 0.08 | 0.00 | 0.03 | 0.04 | 0.00 |
| Variable cost as natural gas | 0.15 | 0.89 | 0.14 | 0.15 | 0.19 | 0.03 |
| Tax on Natural gas | 0.04 | 0.89 | 0.04 | 0.04 | 0.19 | 0.01 |
| Other Variable Costs | | | 0.28 | | | 0.10 |
| Tax on Other Variable Costs | | | 0.05 | | | 0.02 |
| Check test | | | 3.25 | | | 3.00 |
| Tax Discount | | | | | | |
| Tax discount on feedstock | 0% | | 0.00 | | | 0.00 |
| Tax discount on electricity | 0% | | 0.00 | | | 0.00 |
| Tax Discount on Hydrogen | 0% | | 0.00 | | | 0.00 |
| Tax discount on natural gas | 0% | | 0.00 | | | 0.00 |
| Tax Discount on Other Variable Costs | 0% | | 0.00 | | | 0.00 |

| | | | |
|---|---|---|---|
| Total Variable Cost | | 3.25 | 3.00 |
| Fixed Cost manpower | | 0.003 | 0.001 |
| Fixed Cost Recurring maintenance | | 0.02 | 0.02 |
| Other Fixed Costs | | 0.02 | 0.02 |
| Total Fixed Cost | | 0.04 | 0.04 |
| Capital Cost | | 0.02 | 0.02 |
| Revenue SAF | | 0.003 | 0.001 |
| Revenue by  Products | | 0.02 | 0.02 |
| Premium due to GHG reduction | | 0,02 | 0.02 |
| Premium Price for Renewable by products | | 0.04 | 0.04 |
| Premium Price for SAF | | 0.02 | 0.02 |
| Total revenue | | 0.003 | 0.001 |
| Contribution Margin | | 0.02 | 0.02 |
| Net Margin | | 0.02 | 0.02 |
| Depreciation | | 0.04 | 0.04 |
| Revenue taxes | | | |
| Revenue tax discount | 0% | | |
| Free Cash Flow | | -1.02 | -0.68 |

## 9. Conclusion

At current market prices, both HEFA and ATJ routes generate SAF at costs above regular jet fuel, generating negative economic results, even when by-products are accounted for. This indicates that incentive mechanisms would be required in order to create economic viability to the SAF industry and for reaching legal goals.

Stakeholders from the industry interviewed in this work consider that special tax regimes, a regulated carbon market, capital grants for investment and long-term special price agreements are elements that could incentivize the development of the SAF industry.

Based on average prices between 2021-2024, the impact of different incentives is analyzed from an economic perspective. Individually, none of the incentives studied generate positive cash generation for SAF production. Scenarios grouping the different incentives were simulated and the results show that in certain circumstances there is positive cash generation in SAF production along with by-products. Those scenarios can indicate to policy makers the impact of potential public policies necessary to incentivize the SAF industry along with its costs and benefits. For these scenarios, a reverse discounted cash flow analysis was performed, indicating the maximum affordable capital investment supported by both HEFA and ATJ routes in different scenarios. The figures, when negative, indicate that a capital grant would be necessary to pay the investment. When positive, the figures indicate that the cash generation will eventually return the investment in the considered project time frame of 20 years. Moreover, these figures can be compared with the announced investments and can serve as indications of the total capital expenditure necessary to create the production capacity required to meet Brazilian legal goals.

The simulations and economic analysis highlight the relevance of the by-products, in this study valued as per nafta price, for the economic sustainability of SAF production. The impact of these streams and the premiums associated to it and to SAF itself indicate possible ranges of long-term commercial agreements between SAF producers, SAF consumers and the chemical industry which could be a relevant player for the consumption of the renewable by-products generated.

Besides there is still not a regulated carbon market in Brazil the simulations indicate that a premium based on GHG offset has a positive impact in the SAF attractiveness.

From a demand perspective, the volumes required to reach Brazilian goals set by legislation until 2037 range from 1864 kt/y assuming lowest CI and ProBIOQAV goal to 7274 kt/y, considering highest CI and CORSIA goal. Given the magnitude of volumes and the associated investment required, more detailed studies considering location, feedstock availability, distance to consumers, potential supply chain bottlenecks, eligibility to international markets among other elements would be required as a support for elaboration of public policies and investment decision making.

Based on the references [27] and [28] (see table 6), the above-mentioned volumes of SAF would also lead to a byproducts production volume between 1500 kt/y and 6474 kt/y of biogenic by-products, considered in this study as bio-nafta, but in fact, containing a variety of biogenic hydrocarbons. The relevant volume, its importance for the economic viability of SAF production and the potential application of these products in different industries such as biofuels, chemical and petrochemical would also justify a more detailed analysis by the academy and the industrial stakeholders.

From a capital perspective, taking into account the announced investments and the demand estimated to achieve CORSIA and/or ProBioQAV goals, it is possible to estimate the total investment necessary to produce the above-mentioned volumes of SAF in the order of magnitude between R$ 23 and R$ 72 billion or US$ 4.4 and US$ 13.8 Billion, from 2025 to 2037, only in production plants without considering additional facilities for feedstock production, infrastructure and transportation systems. The effect of this amount of investment in the Brazilian economy, along with the evaluation of alternatives for funding it is also a potential area for further investigation and also a perspective to be considered by policy makers, finance institutions and industry stakeholders.

Considering the challenges identified, a deeper evaluation of the feasibility of the potential incentives, especially a specific and special tax regime in the context of a regulatory sandbox framework along with capital grants schemes would be a suggestion for future studies, in order to assess the total cost for the implementation of the legal goals established by the new regulation. Such investigations could provide evaluation of the potential impacts of SAF introduction along the value chains directly affected, such as the soybeans, ethanol, fuels, chemicals, air transport, capital goods among other industries, but also the impacts to the society in general, such as jobs and income generation, impact in GDP and taxes among others.

---